\documentclass[11pt,aps,amsmath,amssymb,nofootinbib,notitlepage,longbibliography]{revtex4-1}
\usepackage{amsmath,amssymb}
\usepackage{slashed}
\usepackage{graphicx}
\usepackage{bm}
\usepackage[top=1.0in,bottom=1.0in,left=1.0in,right=1.0in]{geometry}
\usepackage[colorlinks,linkcolor=blue,citecolor=blue]{hyperref}

\newcommand{\be}{\begin{equation}}
\newcommand{\ee}{\end{equation}}
\newcommand{\bea}{\begin{eqnarray}}
\newcommand{\eea}{\end{eqnarray}}
\newcommand{\ba}{\begin{eqnarray}}
\newcommand{\ea}{\end{eqnarray}}

\newcommand{\Tr}{\mbox{Tr}\;}

\begin{document}

\title{Holographic charm and bottom pentaquarks III\\
Excitations through photo-production of heavy mesons}


\author{Yizhuang Liu}
\email{yizhuang.liu@uj.edu.pl}
\affiliation{Institute of Theoretical Physics,
Jagiellonian University, 30-348 Kraków, Poland}

\author{Kiminad A. Mamo}
\email{kmamo@anl.gov}
\affiliation{Physics Division, Argonne National Laboratory, Argonne, Illinois 60439, USA
}

\author{Maciej A. Nowak}
\email{maciej.a.nowak@uj.edu.pl}
\affiliation{Institute of Theoretical Physics and Mark Kac Center for Complex Systems Research,
Jagiellonian University, 30-348 Kraków, Poland}

\author{Ismail Zahed}
\email{ismail.zahed@stonybrook.edu}
\affiliation{Center for Nuclear Theory, Department of Physics and Astronomy, Stony Brook University, Stony Brook, New York 11794--3800, USA}



\begin{abstract}
We consider the photo-excitation of charm and bottom pentaquarks with the holographic assignments
$[\frac 12\frac 12^-]_{S=0,1}$  and $[\frac 12\frac 32^-]_{S=1}$, in the photo-production
of heavy vector mesons such as charmonia and bottomonia near threshold. We use a Witten
diagram to combine the s-channel photo-excitation of holographic pentaquarks with a massive
t-channel  graviton or tensor glueball exchange, to extract the scattering amplitude for heavy meson
photo-production in the threshold region. The pentaquark signal   is too weak to
be detected at current electron facilities.
\end{abstract}

\maketitle

\section{Introduction}

The recently released data from the  LHCb collaboration show the existence of three narrow pentaquark states
$P_c(4312,4440,4457)$~\cite{Aaij:2019vzc}. This new high statistics data supersedes the previously
reported $P_c(4450)$,   and weakens the status of the previously reported $P_c(4380)$~\cite{Aaij:2015tga}.
This supports the holographic pentaquark construction by two of us where three  degenerate   pentaquarks
$[\frac 12\frac 12^-]_{S=0,1}$  and $[\frac 12\frac 32^-]_{S=1}$ were predicted,
in the triple limit of a large number of colors, a large $^\prime$t Hooft coupling $\lambda$,
and a heavy quark mass~\cite{Liu:2017xzo,Liu:2017frj}.  Similar assignments were subsequently
 made using the molecular construction~\cite{Liu:2019tjn,Xiao:2019aya,Du:2021fmf,Yan:2021nio}.

The three holographic pentaquark states split when spin-orbit interactions  are considered~\cite{Liu:2021tpq,Liu:2021ixf}.
The split  states are not
only close to the masses reported by LHCb, but their  partial decay widths to a variety of open and
hidden channels with heavy charmed mesons are  also close to some  of the  newly reported widths by LHCb~\cite{Aaij:2019vzc}.
The holographic results extend to three new bottom pentaquark states.
The newly reported and narrow $P_c(4337)$  at  3-sigma significance~\cite{LHCb:2021chn},
is not supported by our  holographic analysis of the low-lying
 pentaquark states. The even and odd parity excited holographic pentaquark states $P_c^*$ are heavier.


For completeness, we note that pentaquark states with hidden charm were suggested in~\cite{Voloshin:1976ap,Karliner:2015ina}  and analyzed by many
~\cite{Lebed:2016hpi,Esposito:2016noz,Olsen:2017bmm,Guo:2017jvc,Karliner:2017qhf} (and references therein).
Given the closeness of the reported charmed pentaquarks to hadronic thresholds, most analyses suggest a
molecular form~\cite{Chen:2015loa,Rossi:2016szw,Eides:2017xnt,Guo:2019fdo,Liu:2019tjn,
Chen:2019bip,Chen:2019asm,Eides:2019tgv}. Alternative constructions suggest  solitonic molecules~\cite{Scoccola:2015nia},
string based molecules~\cite{Rossi:2016szw,Sonnenschein:2016ibx}, and light cone based quarkonia~\cite{Dosch:2015bca}.

The holographic charm (bottom) pentaquark states are  bound topological molecules  with hidden charm (bottom),
essentially $[D,D^*]$ ($[B,B^*]$) mesons bound to a topological baryon  at the boundary. They are free of
some of the ambiguities related to the choice of meson exchanges and  form factors~\cite{Voloshin:1976ap,Eides:2017xnt,Lin:2019qiv}.
The dual of the form factors is  the instanton core which is  fixed by gauge-gravity (meson-glueball) interactions in bulk.
 The dual of the meson exchanges are bulk light-light and heavy-light  exchanges warped by the D-brane geometry,
and  the dual of the core gluon exchanges  are bulk gravity and gravitons,
 in conformity with  confinement, chiral symmetry, heavy quark symmetry, and vector dominance.
 The relativistic amplitudes Reggeize with proper crossing symmetries, as expected from a string based formulation. The construction
 involves very few parameters (the flavor brane tension, the Kaluza-Klein (KK)-scale and the mean heavy-light meson mass).

A more detailed  way to understand the structure of the holographic pentaquarks,
is through their strong two-body decay channels~\cite{Liu:2021ixf}, or even better by directly producing them using photo-production
of charmonium (bottomonium) on the proton in the threshold region~\cite{Wang:2015jsa,Kubarovsky:2015aaa,Karliner:2015voa} as currently
pursued at JLab~\cite{Meziani:2016lhg}. Precision data are
required to disentangle the s-channel production from the dominant t-channel contribution, in the nearly diffractive regime.
The purpose of this paper is to address this issue, solely in the holographic construction whereby
three degenerate pentaquark states emerge  with specific spin-isospin-parity assignments and interactions.


The organization of this paper is as follows: in section~\ref{VERTEX} we characterize the holographic transition form factor for
the photo-process $\gamma+ p\rightarrow X$ for the three holographic pentaquarks with  $X=P_{c,b}$.
In the heavy quark limit, it is Pauli-like with the transition magnetic moment $\mu_X/\mu_N\sim (m_N/m_X)^{\frac 32}$. The form factor for the decay process
$X\rightarrow V+p$ with $V=J/\Psi, \Upsilon$ is also discussed. In section~\ref{CROSS} we construct the s-channel photo-production of charmonium
(bottomium) by photo-excitation of a pentaquark state using a Witten diagram. We then  assess its contribution
to the differential cross section, including the t-channel contribution from the graviton or tensor glueball exchange. In section~\ref{NUMERICS}
we give a numerical assessment of each contribution to the differential and total cross sections  in the threshold region.
Our conclusions are in section~\ref{CONCLUSION}.

\section{Transition form factors}~\label{VERTEX}

\begin{figure}[!htb]
\includegraphics[height=12cm,width=10cm]{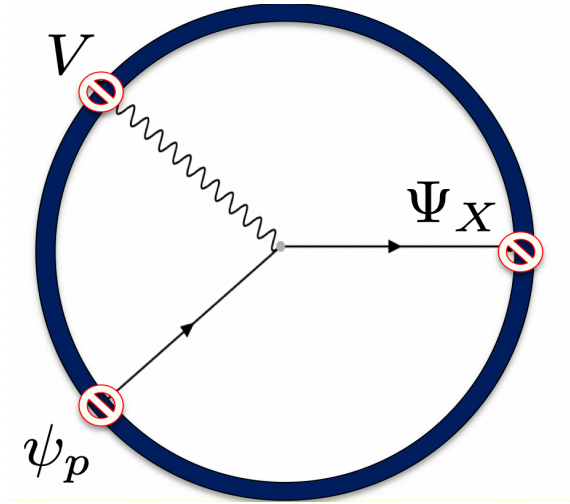}
 \caption{Witten diagram for the U(1) transition vertex $V+\psi_p\rightarrow \Psi_X$.}
  \label{fig_three}
\end{figure}

The transition form factors for pentaquark production $\gamma +p\rightarrow X$ and decay $X\rightarrow V+p$,
were derived using the holographic construction in~\cite{Liu:2021ixf} to which we refer for completeness.
In short, the transition form factor is given by the 3-point Witten diagram  illustrated in Fig.~\ref{fig_three}.
It is composed of the U(1) gauge field in bulk, coupled with a Dirac fermion (probed by the nucleon
at the boundary) and an additional Dirac fermion probed by a pentaquark in the $[\frac 12\frac 12^-]$
representation. The coupling to a pentaquark in the $[\frac 12\frac 32^-]$  representation will follow by
symmetry below.

\subsection{Sakai-Sugimoto model}

In  the Sakai-Sugimoto model~\cite{Sakai:2004cn} the holographic pentaquarks are realized as  topological bound states of a core instanton
with a degenerate  doublet of heavy-light mesons $[D,D^*]$ or $[B,B^*]$ that transmute to spin-$\frac 12$ fermions~\cite{Liu:2017xzo}.
Their decay widths were recently
analyzed  in~\cite{Liu:2021ixf}.
The masses and total widths of the three
charm pentaquark states are fixed to those reported empirically by LHCb~\cite{Aaij:2019vzc}

 \bea
\label{LHCBMW}
m_{P_c}&=&4311.9\pm 0.7\,{\rm MeV}\qquad \Gamma_{P_c}=9.8\pm 2.7\,{\rm MeV}\nonumber\\
m_{P_c}&=&4440.3\pm 1.3\,{\rm MeV}\qquad \Gamma_{P_c}=20.6\pm 4.9\,{\rm MeV}\nonumber\\
m_{P_c}&=&4457.3\pm 0.6\,{\rm MeV}\qquad \Gamma_{P_c}=6.4\pm 2.0\,{\rm MeV}
\eea
For the un-observed  bottom pentaquarks, the theoretical masses and widths estimates from the more recent
holographic analyses~\cite{Liu:2021tpq,Liu:2021ixf}

  \bea
\label{LHCBMW}
m_{P_b}&=&11155\,{\rm MeV}\qquad \Gamma_{P_b}=17.87\,\Gamma=71\pm 18\,{\rm MeV}\nonumber\\
m_{P_b}&=&11163\,{\rm MeV}\qquad \Gamma_{P_b}=10.76\,\Gamma=43\pm 11\,{\rm MeV}\nonumber\\
m_{P_b}&=&11167\,{\rm MeV}\qquad \Gamma_{P_b}=8.15\,\Gamma=33\pm 8\,{\rm MeV}
\eea
 with $\Gamma=4\pm 1$ MeV fixed empirically. The holographic transition couplings will be detailed below.

In the heavy quark limit, the U(1) transition form factor $X=P_{c,b}\rightarrow V+p$  was obtained in the form~\cite{Liu:2021ixf}

\bea
\label{1}
\langle P|{\vec {\mathbb J}}({\vec x-\vec X})|X\rangle=G(\vec P)\,(i\vec{P}\times\overline{v}_{\bar Q}\vec{\sigma}u_{Q}) (2\pi)^3\delta^3(P'-P)\ ,
\eea
with

\begin{align}
\label{2}
G(\vec{P})=\lambda \sqrt{\frac{m_N}{m_{X}}}\sum_{n=\rm odd}
\int dZ \bigg(\Psi_{10}(Z)\frac{\varphi_n(Z)}{\sqrt{\kappa}}\Psi_{10}(Z)\bigg) \frac{g_n}{E^2-\vec{P}^2-\tilde m_n^2}
\end{align}
where $\kappa$ is the flavor brane tension.
The modular wavefunctions $\Psi_{ln_\rho}(Z)$ for the nucleon  and the  pentaquark states are
similar in the ground state

\be
\label{3}
\Psi_{10}(Z)=\bigg(\frac{2M_0}{\sqrt 6 \pi}\bigg)^{\frac 14}\,e^{-\frac{M_0}{\sqrt 6} Z^2}
\ee
with $M_0=8\pi^2 \kappa$ the bulk instanton bare mass in units of $M_{KK}$.  We recall that in the bound state approach, the negative intrinsic parity of
the pentaquark is due to the attached D-meson (B-meson) multiplet to the instanton core. This is  manifest in  the matrix element $\overline{u}_Q\vec{\sigma}v_{\bar Q}$
in (\ref{1}) after spin-statistics  transmutation~\cite{Liu:2017xzo}. The attachment is BPS in leading order.
The modular wavefunction (\ref{3}) is parity even, while the emerging attachment $Q\bar Q$ is parity odd.

The sum in (\ref{2})
is over the  tower of $1^{--}$ vector mesons in bulk. In other words, the transition form factor (\ref{2})  is vector-like.
It is sourced by a  bound pentaquark composed of an  instanton core made of light flavors,  with an emerging  heavy $Q\bar Q$ attachment through
transmutation.  The $1^{--}$ vectors satisfy the eigenvalue equation~\cite{Sakai:2004cn}

\begin{eqnarray}
\label{EIGEN}
-h^{-1}(Z) \partial_Z (k(Z)\partial_Z \varphi_{n}(Z)) = \lambda_n \varphi_n(Z),
\end{eqnarray}
with warping functions $h(Z), k(Z)$~\cite{Sakai:2004cn}.
The eigenvalues are related to the masses of the vector mesons $\tilde m_n=\lambda_n$ in units of $M_{KK}$,
which do not Reggeize. The vector meson decay
constants are $g_{n}\sim 2\kappa (Z\varphi_{2n-1}(Z))$ at asymptotic $Z$.
The eigenvalues are normalized so that $\varphi_n(Z)\sim 1/\sqrt{\kappa}$ with $\kappa$ the brane tension parameter, and
 their  parity is $\varphi_n(-Z)=(-1)^{n+1}\varphi_n(Z)$ according to the mode number.
 The $1^{--}$  trajectory carries odd-n, with the lowest rho-meson  mass $m_1=0.67$.

The boundary gauge invariant, relativistic and local coupling  between the nucleon  and the pentaquark   corresponding to (\ref{1}-\ref{2})  is

\begin{align}
\label{REL}
\frac{e\alpha_X }{m_{X}}\bar \Psi(x)\gamma_{\mu\nu}F^{\mu\nu}(x)\psi(x) \ ,
\end{align}
with matching {positive} parity between the bulk modular wavefunction $\Psi_{10}(Z)$ in (\ref{3}) and the boundary field $\Psi(x)$
with spin-isospin assignment $[\frac 12\frac 12^+]$.
 In Appendix, we show that (\ref{REL})  reproduces the pentaquark
decay width $X\rightarrow V+p$ following from (\ref{1}-\ref{2}) as recently derived in the Sakai-Sugimoto model for heavy baryons~\cite{Liu:2021ixf}.
It is Pauli-like with a transition magnetic moment

\begin{align}
\label{MUX}
\frac{\mu_{X}}{\mu_N}=\frac{e\alpha_X }{\mu_N m_{X}}\rightarrow 4\eta_X\frac{m_N^2}{m_\rho^2}
\end{align}
in nuclear Bohr magneton $\mu_N=e /2m_N$. The  last identity follows from mapping to  the soft wall, and
using (\ref{PARAX}) below.
For a charm pentaquark mass  $m_X\sim 4440$ MeV we have $\mu_{4440}/\mu_N\sim 0.78$, with
 the transition coupling  $|\eta_{4440}|=0.134$ using (\ref{ETAX}).
The transition magnetic moment is smaller for bottom pentaquarks since
$\mu_{P_b}/\mu_{P_c}=(m_{P_c}/m_{P_b})^{\frac 32}$ following from (\ref{MUX}) and (\ref{CONSTANT}).

\subsection{Soft-Wall model}

In the Sakai-Sugimoto model the resummed $1^{--}$ spectrum does not Reggeize, as we noted earlier. As a result,
the ensuing U(1) form factors do not obey the hard scattering rules asymptotically.
To fix this we will use instead the soft wall model which Reggeizes, while
maintaining the same Pauli-like form factors for all holographic pentaquark states
in the heavy quark limit by matching.

With this in mind, the bulk interaction vertex between the Dirac fermion $\psi$  and the pentaquark field $\Psi$ is

\begin{align}
\label{PAULI1}
\eta_X \int dzd^4x \sqrt{|g(z)|} \,e^{-\phi_N(z)}\, \sum_{\xi=1,2}\bar \Psi_{\xi}e^{M}_Ae^{N}_B \sigma^{AB} F_{MN} \psi_{\xi} \ ,
\end{align}
with $g_{MN}(z)=(R^2/z^2)(+,-,-,-,-)$ the mostly negative  AdS$_5$ metric, $e_A^M$ its associated vierbeins,
$\sigma^{AB}=\frac i2 [\Gamma^A, \Gamma^B]$, and
$\phi_N(z)=\tilde\kappa_N^2 z^2$ the dilaton in the bulk filling baryonic brane in
the soft wall approximation. Note that the $^\prime$pentaquark$^\prime$ bulk field $\Psi (x,z)$ is the
 the  analogue of the modular field (\ref{3}) with positive parity. This is important in describing the analogue of the
 vector form factor (\ref{2}) in the soft wall construction with manifest Regge behavior.
The labels $\xi=1,2$ refer to the chiralities of the fermionic operators inserted at the boundary~\cite{Mamo:2021cle} and

\begin{align}\label{eq:couple}
\frac{\eta_X}{{\tilde\kappa_N}  }= \frac{\alpha_X}{m_{X}}
\end{align}
is fixed by comparing the two holographic constructions.
The initial and final state wave functions for the nucleon  (pentaquark)  states  will be defined as

\bea
\psi(z)&=&\psi_L(z)+\psi_R(z)\nonumber\\
\psi_{L,R}(z)&=&e^{-iP\cdot x}\frac{1}{2}(1\mp \gamma^5)u_s(P) \frac{z^2}{R^2}\tilde f_{L,R}(z) \ ,
\eea
where the left and right modes satisfy the equation of motion

\begin{align}
\bigg[-\partial_z^2+\tilde\kappa_N^2z^2+2\tilde\kappa_N^2(M_N\mp\frac{1}{2})+\frac{M_N(M_N\pm1)}{z^2}\bigg]\tilde f_{L,R}(z)=P^2\tilde f_{L,R}(z) \ .
\end{align}
with $M_N=\Delta_N-2=\tau_N-\frac 32$ the anomalous dimension of the nucleon (pentaquark)  source with twist $\tau_N$.
The mass spectrum Reggeizes radially

\begin{align}
P^2=m_n^2=4\tilde\kappa_N^2\bigg(n+M_N+\frac{1}{2}\bigg) \ .
\end{align}
with the normalized states

\begin{align}
&\tilde f^n_{L}(z)=\sqrt{\frac{2\Gamma(n+1)}{\Gamma(n+M_N+\frac{3}{2})}}\tilde\kappa_N^{M_N+\frac{3}{2}}z^{M_N+1}L_n^{M_N+\frac{1}{2}}(\tilde\kappa_N^2z^2)
\equiv C_{L}(M_N,n)\,\kappa^{M_N+\frac{3}{2}}z^{M_N+1}L_n^{M_N+\frac{1}{2}}(\tilde\kappa_N^2z^2)
\nonumber\\
&\tilde f^n_{R}(z)=\sqrt{\frac{2\Gamma(n+1)}{\Gamma(n+M_N+\frac{1}{2})}}\tilde\kappa_N^{M_N+\frac{1}{2}}z^{M_N}L_n^{M_N-\frac{1}{2}}(\tilde\kappa_N^2z^2)
\equiv C_{R}(M_N,n)\tilde\kappa_N^{M_N+\frac{1}{2}}z^{M_N}L_n^{M_N-\frac{1}{2}}(\tilde\kappa_N^2z^2) \ .\nonumber\\
\end{align}
For the proton, one has $n=0$ and the anomalous dimension is
$M_N \equiv\tau_N-\frac{3}{2}$. For pentaquark states, one can choose a generic anomalous dimension $M_{X}$ and radial
quantum number $n=n_X$, assuming a universal baryonic slope $\tilde\kappa_X=\tilde\kappa_N$. The twist $\tau_X$ and the
anomalous dimensions $M_{X}$ are fixed below.
Note that for  ground state ($n=0$) the bulk fermionic modes simplify
\bea
\label{PLRZ}
\psi_L(z,\tau_N)&=& z^{\Delta_N}\times\tilde{\psi}_L(z,\tau_N)=n_L(\tilde\kappa_N,\tau_N) \times\tilde\kappa_N^{2(\tau_N-1)}z^{2(\tau_N-1)}\times z^{\Delta_N}\,,\nonumber\\
\psi_R(z,\tau_N)&=&z^{\Delta_N}\times\tilde{\psi}_R(z,\tau_N)=n_R(\tilde\kappa_N,\tau_N) \times \tilde\kappa_N^{2(\tau_N-1)-1}z^{2(\tau_N-1)-1}\times z^{\Delta_N}\,,\nonumber\\
\eea
with the normalization factors

\bea
n_L(\tilde\kappa_N,\tau_N)&=&\frac{1}{\tilde\kappa_N^{\tau -2}}\times\sqrt{\frac{2}{\Gamma(\tau_N)}}\,,\nonumber\\
n_R(\tilde\kappa_N,\tau_N)&=&n_L(\tilde\kappa_N,\tau)\times\sqrt{\tau_N-1}\,.
\eea

\subsection{Transition form factor: $\gamma+ p\rightarrow P_c$}

The re-summed U(1) bulk-to-boundary propagator in the soft-wall model is~\cite{Mamo:2019mka,Mamo:2021cle},

\bea
\label{VQZ}
&&\mathcal{V}(Q,z,\tilde{\kappa}_{V})
=\tilde\kappa_V^2z^2 \,\Gamma(1+a_Q)\,\,{\cal U} (1+a_Q; 2 ; \tilde\kappa_V^2z^2)\nonumber\\
&&=\tilde\kappa_V^2z^2\int_{0}^{1}\frac{dx}{(1-x)^2}x^{a_Q}{\rm exp}\Big(-\tilde\kappa_V^2z^2\frac{x}{1-x}\Big)
=a_Q\int_{0}^{1} dx\,x^{a_Q-1}{\rm exp}\Big(-\tilde\kappa_V^2z^2\frac{x}{1-x}\Big)\,,\nonumber\\
\eea
with ${\cal U}$ the Kummer function and   $a_Q=Q^2/(4\tilde \kappa_V^2)$. We note the normalizations  ${\cal V}(0,z,\tilde\kappa_V)={\cal V}(Q,0,\tilde\kappa_V)=1$.
As a result, for the Pauli-like coupling (\ref{PAULI1}), the form factor becomes

\bea
&&{\cal W}(Q^2)=\frac{\eta_X}{\tilde\kappa_N} \bar u(P')i(\slashed{\epsilon}\slashed{q}-\slashed{q}\slashed{\epsilon})u(P){\cal I}(n_X,Q^2) \ ,\\
&&{\cal I}(n_X,Q^2)=\frac{1}{2}\bigg({\cal I}_{LR}(n_X,Q^2)+{\cal I}_{RL}(n_X,Q^2)\bigg) \ ,
\eea
where the scalar form factors ${\cal I}_{LR}$ and ${\cal I}_{RL}$ read

\bea
&&{\cal I}_{LR}(n_X,Q^2)=\frac{1}{2}C_L(M_X,n_X)C_R(M_N,0) \nonumber \\ \times &&\int_{0}^{\infty} dw w^{\frac{M_X+M_N+3}{2}}e^{-w} L^{M_X
+\frac{1}{2}}_{n_X}(w){\cal U}\big(1+a_Q;2,w\big)\Gamma\big(1+a_Q\big) \ ,\nonumber \\
&&{\cal I}_{RL}(n_X,Q^2)=\frac{1}{2}C_R(M_X,n_X)C_L(M_N,0) \nonumber \\ \times &&\int_{0}^{\infty} dw w^{\frac{M_X+M_N+3}{2}}e^{-w} L^{M_X-\frac{1}{2}}_{n_X}(w)
{\cal U}\big(1+a_Q;2,w\big)\Gamma\big(1+a_Q\big) \ .
\eea
Notice that for identical anomalous dimensions $M_X=M_N$, the above integrals can be analytically evaluated.

\subsection{Transition form factor: $P_c\rightarrow  V  +p$}

When the $U(1)_{V}$ probing source goes on-shell a vector meson $V=J/\Psi, \Upsilon$ is produced.
Using the LSZ reduction, we first note that


\begin{align}
w\,{\cal U}\big(1+a_Q;2,w\big)\Gamma\big(1+a_Q\big)=\sum_{n}\frac{\phi_n(z)F_n}{Q^2+\tilde m_n^2} \rightarrow \phi_n(z) \ ,
\end{align}
with the Reggeized meson spectrum $\tilde m_n^2=2\tilde\kappa_V^2(n+1+C_V)$ shifted by $C_V=M_V^2/2\tilde\kappa^2-1$ and

\begin{align}
\phi_n(z)= \sqrt{\frac{2}{n+1}}\tilde\kappa_V^2z^2 L_n^1(\tilde \kappa_V^2z^2) \ ,
\end{align}
the normalized wave functions for vector mesons $V$ in bulk. Therefore, the form factor for $X\rightarrow V+p$ is

\bea
\label{WMN2}
&&{\cal W}(-m_n^2)=\frac{\eta_X}{\tilde\kappa_N} \bar u(P')i(\slashed{\epsilon}\slashed{q}-\slashed{q}\slashed{\epsilon})u(P){\cal I}(n_X, -m_n^2) \ ,\nonumber\\
&&{\cal I}(n_X,-m_n^2)=\frac{1}{2}\bigg({\cal I}_{LR}(n_X, -m_n^2)+{\cal I}_{RL}(n_X, -m_n^2)\bigg) \ ,
\eea
with

\bea
&{\cal I}_{LR}(n_X,-m_n^2)=\frac{1}{2}C_L(M_X,n_X)C_R(M_N,0)\sqrt{\frac{2}{n+1}}
\int_{0}^{\infty} dw w^{\frac{M_X+M_N+3}{2}}e^{-w} L^{M_X+\frac{1}{2}}_{n_X}(w)L_n^1(w)\ , \nonumber\\\nonumber\\
&{\cal I}_{RL}(n_X,-m_n^2)=\frac{1}{2}C_R(M_X,n_X)C_L(M_N,0)\sqrt{\frac{2}{n+1}}
\int_{0}^{\infty} dw w^{\frac{M_X+M_N+3}{2}}e^{-w} L^{M_X-\frac{1}{2}}_{n_X}(w)L_n^1(w) \ .\nonumber\\
\eea

\begin{figure}[!htb]
\includegraphics[height=12cm,width=16cm]{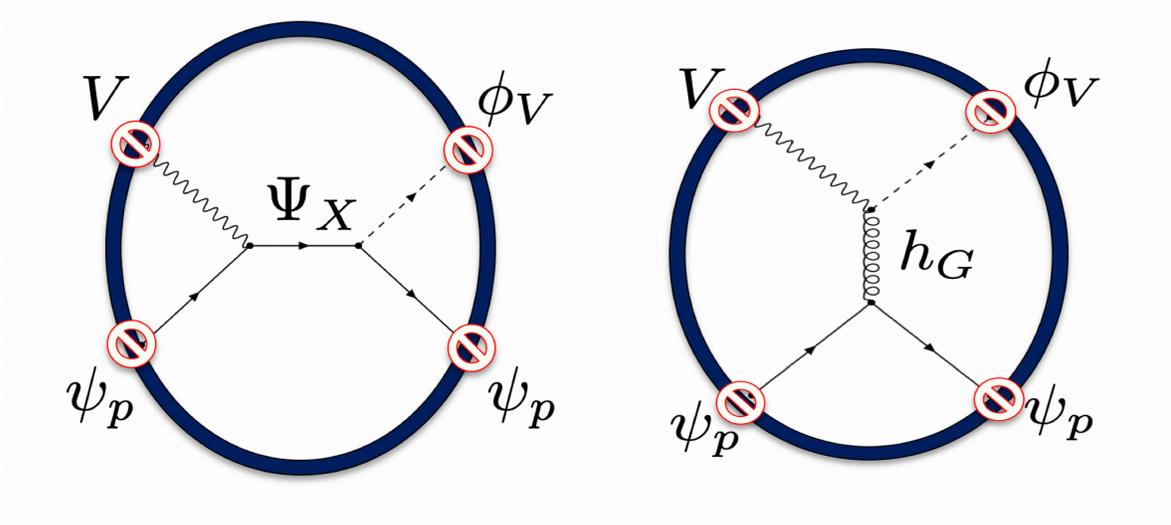}
 \caption{Witten diagram for the photo-production process $V+p\rightarrow p+V$ of a vector meson $V$ off a proton $\psi_p$.
 The pentaquark $\Psi_X$ excitation in the s-channel with no crossing shown (left)  is added to
the  graviton exchange $h_G$  in the t-channel (right).
}
  \label{fig_photo}
\end{figure}

\section{Differential cross section}~\label{CROSS}

In holography, the photo-production of charmonium (bottomonium) near threshold receives s- and t-channel contributions.
The holographic t-channel  contribution in Fig.~\ref{fig_photo} (right)  is known~\cite{Mamo:2019mka,Mamo:2021tzd}.
It  is dominated by the graviton exchange (tensor glueball)
near threshold as detailed in~\cite{Mamo:2019mka}

\bea
\label{DIFFG}
\left(\frac{d\sigma}{dt}\right)_G
=\frac{{\cal N}^2e^2}{64\pi (s-m_N^2)^2}\times\frac{A^2(Q)}{A^2(0)}\times \bigg(2-\frac{t}{2m_N^2}\bigg)\,F(s,t)
\eea
with the gravitational form factor

\be
\label{AG}
A(Q)=A(0)\,(a_Q+1)\bigg(-\left(1+a_Q+2a_Q^2\right)+2\left({a_Q}+2{a^3_Q}\right)\Phi(-1,1,a_Q)\bigg) \,,
\ee
$\Phi(-1,1,a_Q)$ refers to the Lerch $\Phi$  function, with $a_Q={Q^2}/{8\tilde\kappa_V^2}$  and $Q^2=-t$. It  is remarkably dipole-like.
The extra kinematical factor
$F(s,t)$ can be found in~\cite{Mamo:2019mka}. It scales like $s^4$ at large $s$ but otherwise varies weakly with $t$ near treshold.

\subsection{Spin $\frac 12$}

The s-channel contribution is new, and will be restricted to the
possible  photo-excitation of a pentaquark as a possible threshold enhancement. More specifically, the s-channel $[\frac 12\frac 12^-]_{S=0,1}$ pentaquark contribution
in Fig.~\ref{fig_photo} (left) plus crossing give

\begin{align}
{\cal M}_{\frac{1}{2}}=\sum_{n_X}&\frac{[\frac{\eta_X}{\tilde\kappa_N}{\cal I}(0,0)][\frac{\eta_X}{\tilde\kappa_N}{\cal I}(0, -m_V^2)]}{s-m_{n_X}^2+im_{n_X}\Gamma_{n_X}}\nonumber \\ &\times \bar u(p',S')\left(\slashed{k}'\slashed{\epsilon}^{\star}-\slashed{\epsilon}^{\star}\slashed{k}'\right)(\slashed{p+k}+m_{n_X})\left(\slashed{\epsilon}\slashed{k}-\slashed{k}\slashed{\epsilon}\right)u(p,S)+{\rm cross}
\end{align}
Keeping only the  ground states $n_X=0$ contribution, then we approximately have

\begin{align}
{\cal M}_{\frac{1}{2}}\approx e^2\,\frac{[\frac{\eta_X}{\tilde\kappa_N}{\cal I}(0,0)][\frac{\eta_X}{\tilde\kappa_N}{\cal I}(0, -m_V^2)]}{s-m_X^2+im_X\Gamma_{X}}{\cal A}_{\frac{1}{2}}(S^\prime,S)+{\rm cross} \ ,
\end{align}
with $\eta_X=\alpha_X/m_X$, $m_X=m_{n_X=0}$ and  $\Gamma_X=\Gamma_{n_X=0}$  and

\begin{align}
{\cal A}_{\frac{1}{2}}(S',S) =\bar u(p',S')\left(\slashed{k}'\slashed{\epsilon}^{\star}-\slashed{\epsilon}^{\star}\slashed{k}'\right)(\slashed{p+k}+m_{n_X})\left(\slashed{\epsilon}\slashed{k}-\slashed{k}\slashed{\epsilon}\right)u(p,S)
\end{align}
is the generic-spin structure for the Pauli-exchange. For the near-threshold production, $p\approx \frac{m_X}{2}(1,1,0,0)$ and $k\approx \frac{m_X}{2}(1,-1,0,0)$ are close to the light-cone, while $\epsilon$ and $\epsilon'$ are transversally polarized. The crossing contribution is expected to be small for a far off-shell $X=P_{c,b}$. It will be ignored. Hence

\bea
&&\frac{1}{4}\sum_{S,S^\prime}|{\cal A}_{\frac{1}{2}}(S^\prime,S)|^2=\\
&&4\,{\rm Tr}\bigg[(\slashed{p}+\slashed{k}+m_X)(3m_Nk^2+k^2\slashed{p}-4k\cdot p\slashed{k})(\slashed{p}
+\slashed{k}+m_X)(3m_N k^{\prime 2}+(k^{\prime 2}\slashed{p}^{\prime}-4k^\prime\cdot p^\prime{\slashed{k}}^\prime)\bigg]\nonumber
\eea
after summing over the  polarizations and spins. Here we have used the identity
$$(\gamma^{\mu}\slashed{k}-\slashed{k}\gamma^{\mu})(\slashed{p}+m_N)(\slashed{k}\gamma_{\mu}-\gamma_{\mu}\slashed{k})=12k^2m_N+4k^2\slashed{p}-16k\cdot p\slashed{k}.$$
In general, the trace can be evaluated using $k^2=-Q^2$ for the electro-production process,
and the standard relation $s+t+u=2m_N^2+m_X^2-Q^2$

\begin{align}\label{eq:trace}
& {\cal A}_{\frac{1}{2}}(s,t; Q^2)\equiv \frac{1}{4}\sum_{S,S^\prime}|{\cal A}_{\frac{1}{2}}(S^\prime,S)|^2 \nonumber \\ =&8 \bigg(-4 m_N^6 \left[3 m_X^2-3 Q^2+4 s+t\right]+m_N^4 \left[m_X^2 \left(-37 Q^2+20 s-2 t\right)+Q^2 (2 t-20 s)+12 s (2 s+t)\right]\nonumber \\ &+m_N^2 \left[m_X^2 \left(Q^2 (t-22 s)+9 Q^4+4 s (t-s)\right)-9 m_X^4 Q^2-4 s \left(Q^2 (t-s)+s (4 s+3 t)\right)\right]\nonumber \\ &+4 m_N^8-m_X^2 s \left[Q^2 (5 s+t)+Q^4+2 s (2 s+t)\right]+m_X^4 Q^2 \left[Q^2+s\right]+2 s^2 \left[Q^2 (2 s+t)+2 s (s+t)\right] \bigg) \ .
\end{align}

The s-channel contribution of $[\frac 12\frac 12^-]_{S=0,1}$ pentaquark state to the differential cross section for photo-production is

\bea
\label{DIFFX}
\left(\frac{d\sigma}{dt}\right)_X
\approx \frac{e^2}{16\pi(s-m_N^2)^2}\,
\bigg|\frac{[\frac{\eta_X}{\tilde\kappa_N}{\cal I}(0,-Q^2)][\frac{\eta_X}{\tilde\kappa_N}{\cal I}(0, -m_V^2)]}{s-m_X^2+im_X\Gamma_{X}}\bigg|^2 {\cal A}_{\frac{1}{2}}(s,t; 0)
\eea
The kinematic factor ${\cal A}(s,t;0)$ is given by Eq.~(\ref{eq:trace}). The vertex form factors are explicitly

\bea
 \frac{\eta_X}{\tilde\kappa_N}\times {\cal I}(0,-Q^2)
&=&\frac{\eta_X}{\tilde\kappa_N}\times\tilde{\kappa}_N\times\int dz\, e^{-\tilde{\kappa}_N^2z^2}\frac{\mathcal{V}(Q,z,\tilde{\kappa}_{V})}{2z^{\tau_N+\tau_X-4}} \Big[\tilde{\psi}_L(z,\tau_X)\tilde{\psi}_R(z,\tau_N)+\tilde{\psi}_R(z,\tau_X)\tilde{\psi}_L(z,\tau_N)\Big]\nonumber\\
&=&\frac{\eta_X}{\tilde\kappa_N}\times\frac{\Gamma (a_Q+1) \left(\sqrt{\tau_N-1}+\sqrt{\tau_X-1}\right) \Gamma \left(\frac{1}{2} (\tau_N +\tau_X)\right) \Gamma \left(\frac{1}{2} (\tau_N +\tau_X +2)\right)}{4\Gamma (a_Q+\tau_X +1) \sqrt{\Gamma (\tau_N) \Gamma (\tau_X)}}\,,\nonumber\\
\frac{\eta_X}{\tilde\kappa_N}\times {\cal I}(0,-m_V^2)
&=&\frac{\eta_X}{\tilde\kappa_N}\times\tilde{\kappa}_N\times\int dz\, e^{-\tilde{\kappa}_N^2z^2}\frac{\sqrt{2}\tilde{\kappa}_{V}^2z^2}{2z^{\tau_N+\tau_X-4}} \Big[\tilde{\psi}_L(z,\tau_X)\tilde{\psi}_R(z,\tau_N)+\tilde{\psi}_R(z,\tau_X)\tilde{\psi}_L(z,\tau_N)\Big]\nonumber\\
&=&\frac{\eta_X}{\tilde\kappa_N}\times\frac{\tilde\kappa_V^2}{\tilde\kappa_N^2}\times\frac{\left(\sqrt{\tau_N-1}+\sqrt{\tau_X-1}\right) \Gamma \left(\frac{1}{2} (\tau_N +\tau_X +2)\right)}{2\sqrt{2}\sqrt{\Gamma (\tau_N) \Gamma (\tau_X)}}\,,\nonumber\\
\eea
with $a_Q=Q^2/4\tilde\kappa^2_V$, after using (\ref{VQZ}) and (\ref{PLRZ}) respectively.
We fix the Regge slopes and anomalous dimensions as follows

\be
\label{PARAX}
4\tilde\kappa_N^2=4\tilde\kappa_\rho^2=m_\rho^2\,,\qquad \tau_N=1+\frac{m_N^2}{m_\rho^2}=2.465\,,\qquad
\tau_X =1+\frac{m_X^2}{m_\rho^2}\,.
\ee
and set $\tilde\kappa_V =\tilde\kappa_{J/\Psi}=1.03784\,\rm{GeV}$
from  the high-energy electroproduction data for $J/\Psi$ as was done in~\cite{Mamo:2019mka,Mamo:2021tzd}.

\subsection{Pauli coupling constant}

In so far, the only unknown parameter in our set-up (for the soft-wall holographic QCD model) is the Pauli coupling $\eta_X$ which, in principle, can be fixed by the experimental value of the branching ratio for ${Br}(X\rightarrow V+p)$ which is lacking. Therefore, we use the theoretically computed $Br(X\rightarrow V+p)$
from the Sakai-Sugimoto holographic QCD model in~\cite{Liu:2021ixf} to fix $\eta_X$.

In the soft-wall, the partial decay width for $X(p_X)\rightarrow V(k^{\prime})+p(p^{\prime})$ is given by

\bea
\label{DIFFGammaVX}
\frac{d\Gamma_{XV}}{d\Omega}&=&\frac{\vert\vec{k}_V\vert}{32\pi^2 m_X^2}\times\frac 12\sum_{{\rm pol}}
\frac 12\sum_{{\rm spin}}
\Bigg|{\cal A}_{X\rightarrow  J/\Psi p} (m_X,m_N,m_V)\Bigg|^2\,,\nonumber\\
&=& \frac{\vert\vec{k}_V\vert}{32\pi^2 m_X^2}\times
\left[\frac{\eta_X}{\tilde\kappa_N}{\cal I}(0,-m_V^2)\right]^2\times {\cal A}_{\Gamma_{XV}}(m_X,m_N,m_V)\nonumber\\
\eea
with $\vec{k}^{\prime}=\vec{k}_V$, the 3-momentum of the emitted vector meson, given by

\be
\label{KINX}
\vert\vec{k}_V\vert = \frac{1}{2m_X}\sqrt{m_X^4-2(m_V^2+m_N^2)m_X^2+(m_V^2-m_N^2)^2}\,,
\ee
and we have defined

\bea
{\cal A}_{\Gamma_{XV}}(m_X,m_N,m_V)&=&\,\frac 12\sum_{{\rm pol}}\frac 12\Tr\Big[\left(\slashed{p}^\prime +m_N\right)\left(\slashed{\epsilon}^\prime\slashed{k}^\prime-\slashed{k}^\prime\slashed{\epsilon}^\prime\right)\left(\slashed{p}_X+m_X\right)\left(\slashed{k}^\prime\slashed{\epsilon}^{\prime *}-\slashed{\epsilon}^{\prime *}\slashed{k}^\prime\right)\Big]\nonumber\\
&=&-12m_V^2m_Nm_X-4m_V^2p\cdot p'+16k'\cdot p k'\cdot p' \, \nonumber\\
&=&-12m_V^2m_Nm_X-2m_V^2(m_N^2+m_X^2-m_V^2)\nonumber\\
&&+4(m_X^2-m_V^2-m_N^2)(m_X^2+m_V^2-m_N^2) \ , 
\eea
where we have used

\be
\sum_{{\rm s=1,2,3}}\epsilon_{s}^{\prime\mu}\epsilon_{s}^{\prime *\nu}=-\eta^{\mu\nu}+\frac{k^{\prime\mu}k^{\prime\nu}}{m_V^2}\,.
\ee

Therefore, we have

\bea
\label{GammaVX}
\Gamma_{XV}&=& \frac{\vert\vec{k}_V\vert}{8\pi m_X^2}\times
\left[\frac{\eta_X}{\tilde\kappa_N}{\cal I}(0,-m_V^2)\right]^2\times {\cal A}_{\Gamma_{XV}}(m_X,m_N,m_V)\,.
\eea
which allows to fix the Pauli coupling $\eta_X$ given the partial widths $\Gamma_{XV}$

\bea
\label{ETAX}
\eta_X^2 &=& \frac{\Gamma_{XV}}{\frac{\vert\vec{k}_V\vert}{8\pi m_X^2}\times
\left[\frac{1}{\tilde\kappa_N}{\cal I}(0,-m_V^2)\right]^2\times {\cal A}_{\Gamma_{XV}}(m_X,m_N,m_V)}\,.
\eea
Note that the photo-decay widths $\Gamma_{X\gamma}$ of the holographic pentaquarks in the process
  $X(p_X)\rightarrow \gamma(k)+p(p)$
are   fixed by similar arguments

\bea
\label{GammagammaX}
\Gamma_{X\gamma}&=& \frac{\vert\vec{k}\vert}{8\pi m_X^2}\times
\left[\frac{\eta_X}{\tilde\kappa_N}{\cal I}(0,-Q^2)\right]^2\times {\cal A}_{\Gamma_{X\gamma}}\left(m_X,m_N,m_V\rightarrow \sqrt{-Q^2}\right)\,.
\eea

The partial widths $\Gamma_{XV}$ were evaluated in the holographic
construction in~\cite{Liu:2021ixf}. More specifically:
For $X=P_c(4440)$ and $V=J/\Psi$, using $\Gamma_{XV}=0.00034\pm 0.00009\,\rm{GeV}$, we find its Pauli coupling to be $\eta_X(4440)=0.134 \pm 0.017 $.
For this decay process, the  branching ratio is ${\it Br}_X=\Gamma_{XV}/\Gamma_X=1.8\%$, which is close to the upper bound ${\it Br}_{X,\rm GlueX}<2\%$ set by GlueX~\cite{GlueX:2019mkq},  and the lower bound ${\it Br}_{X}>0.05\%$ empirically argued in~\cite{Cao:2019kst}.
For $X=P_c(4312)$ and $V=J/\Psi$, using $\Gamma_{XV}=0.000056\pm 0.000014\,\rm{GeV}$, we find its Pauli coupling to be $\eta_X(4312)=0.044 \pm 0.005$,
and the branching ratio is ${\it Br}_X=0.3\%$ within the bounds.
For $X=P_c(4457)$ and $V=J/\Psi$, using $\Gamma_{XV}=0.00017\pm 0.000043\,\rm{GeV}$, we find its Pauli coupling to be $\eta_X(4457)=0.099 \pm 0.012$.
The branching ratio is ${\it Br}_X=0.9\%$ also within the bounds.
The photo-decay widths are  $\Gamma_{X\gamma}(4312)=(3.2\pm 0.8)\times 10^{-8}\,\rm{GeV}$,  $\Gamma_{X\gamma}(4440)=(1.05\pm 0.26)\times 10^{-7}\,\rm{GeV}$, and $\Gamma_{X\gamma}(4457)\approx 0$. These observations
 extend to the newly predicted bottom pentaquark states in~\cite{Liu:2021tpq,Liu:2021ixf}.

\subsection{spin $\frac 32$}

For a pentaquark state
$[\frac 12\frac 32^-]_{S=1}$ the contributions are similar, with the bulk covariantized contribution~\cite{DHoker:2016ncv}
\bea
\label{PAULI2}
 \tilde\eta_X \int dzd^4x \sqrt{|g(z)|} \,e^{-\phi(z)}\, \sum_{\xi=1,2}\bar \psi_{\xi} e^{M}_Ae^{N}_B e_C^L \Gamma^C \sigma^{AB} F_{MN} \Psi_{L,\xi} \ ,
\eea
instead of (\ref{PAULI1}). The  Rarita-Schwinger propagator of the $[\frac 12\frac 32^-]_{S=1}$ pentaquark at the boundary is

\bea
S^{F}_{\mu\nu}=&&\frac{i\left(\slashed{p}+\slashed{k}+m_X\right)}{s-m_{X}^2+im_X\Gamma_{X}}\nonumber\\
&&\times \left[\eta_{\mu\nu}-\frac{1}{3}\gamma_{\mu}\gamma_{\nu}-\frac{1}{15m_X}\left((p+k)_\mu\gamma_\nu + \gamma_\mu (p+k)_\nu\right)-\frac{8}{15}\frac{(p+k)_\mu (p+k)_\nu}{m_X^2}\right]\,.\nonumber\\
\eea
The net contribution to the squared s-channel amplitude without crossing is

 \begin{align}
 \label{S321}
{\cal M}_{\frac{3}{2}}\approx e^2\,\frac{[\frac{\tilde\eta_X}{\tilde\kappa}{\cal I}(0,0)][\frac{\tilde\eta_X}{\tilde\kappa}{\cal I}(0, -m_V^2)]}{s-m_X^2+im_X\Gamma_{X}}{\cal A}_{\frac{3}{2}}(S^\prime,S)+{\rm cross} \ ,
\end{align}
with

\bea
\label{S322}
{\cal A}_{\frac{3}{2}}(S',S)&=&\bar u(p',S')\times\gamma^\mu\times\left(\slashed{k}'\slashed{\epsilon}^{\star}-\slashed{\epsilon}^{\star}\slashed{k}'\right)\times (\slashed{p}+\slashed{k}+m_{X})\nonumber\\
&\times & \left[\eta_{\mu\nu}-\frac{1}{3}\gamma_{\mu}\gamma_{\nu}-\frac{1}{15m_X}\left((p+k)_\mu\gamma_\nu + \gamma_\mu (p+k)_\nu\right)-\frac{8}{15}\frac{(p+k)_\mu (p+k)_\nu}{m_X^2}\right]\nonumber\\
&\times &\left(\slashed{\epsilon}\slashed{k}-\slashed{k}\slashed{\epsilon}\right)\times\gamma^\nu\times u(p,S)\,.\nonumber\\
\eea

Given (\ref{S321}-\ref{S322}), the contribution of the $[\frac 12\frac 32^-]_{S=1}$ pentaquark to the differential cross section follows the same reasoning
as the $[\frac 12\frac 12^-]_{S=0,1}$ ones. However, the detailed construction is not needed.
Indeed, in the hidden channel decay analysis, we have found that all the decay amplitudes ${\cal M}$ are the same, after summing over the initial and final isospin and spin states. The difference in the partial decay width of the $[\frac 12\frac 12^-]_{S=0}$ and $[\frac 12\frac 32^-]_{S=1}$ pentaquarks,
 is caused by the spin averaging factor $\frac{1}{2S+1}$. Therefore, for the
photo-production case where the X-pentaquark is in the intermediate state,
 and where all   spin-isospin are summed over, the final amplitude is  the same  irrespective of  the intrinsic
 spin values $S=0,1$. The kinematical differences introduce by (\ref{S322}) versus (\ref{eq:trace}) are qualitative.

 \begin{figure}[!htb]
\includegraphics[height=5.5cm]{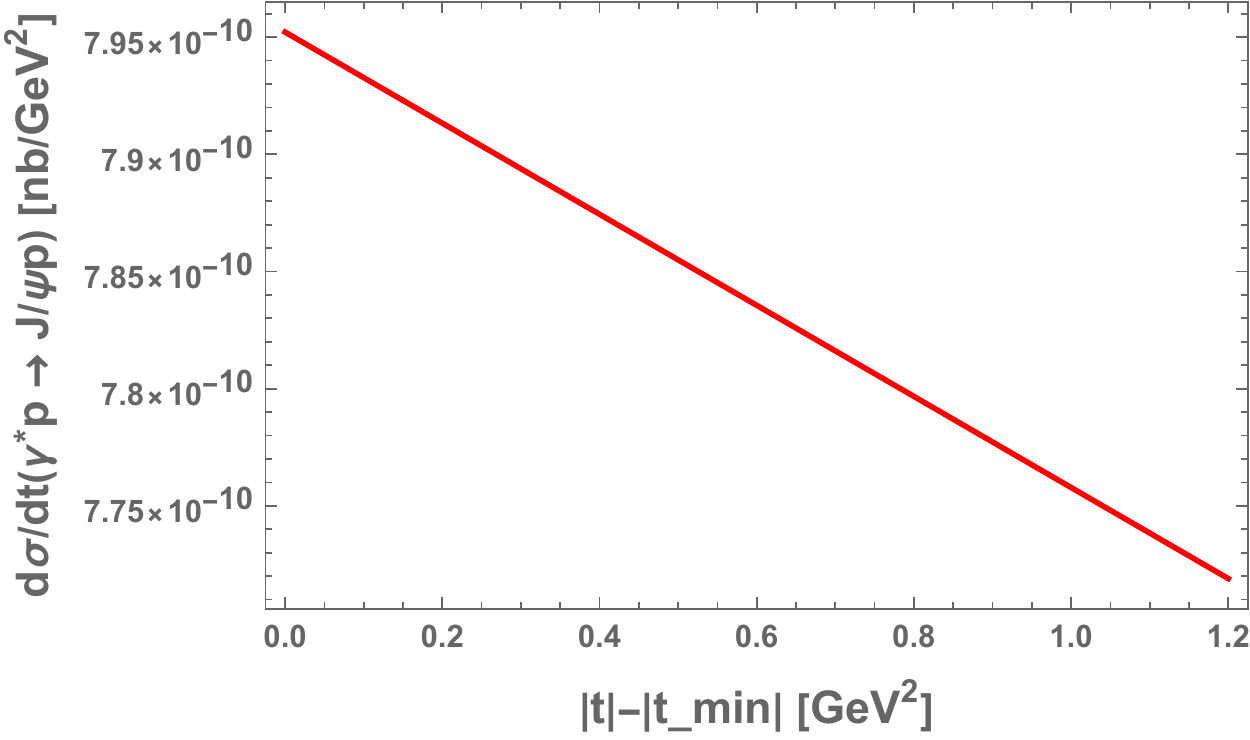}
  \caption{s-channel contribution to the photo-production differential cross section  for $V=J/\Psi$, including all three charm pentaquark contributions.}
  \label{fig_dcsS}
\end{figure}

 \begin{figure}[!htb]
\includegraphics[height=5.5cm]{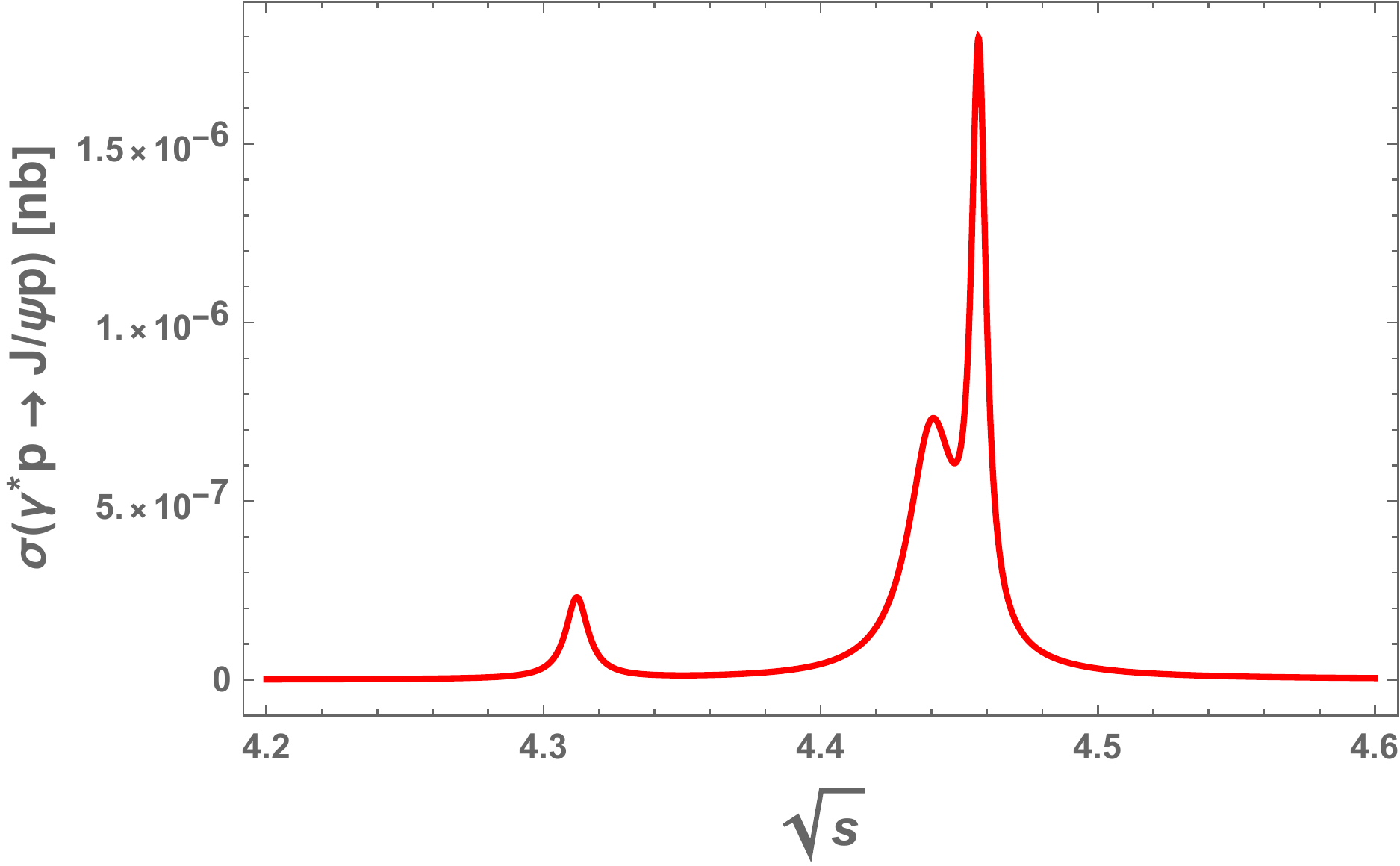}
  \caption{s-channel contribution to the photo-production cross section  for $V=J/\Psi$  versus $\sqrt{s}$,   showing the three charm pentaquarks.}
  \label{fig_totcsS}
\end{figure}

\begin{figure}[!htb]
\includegraphics[height=5.5cm]{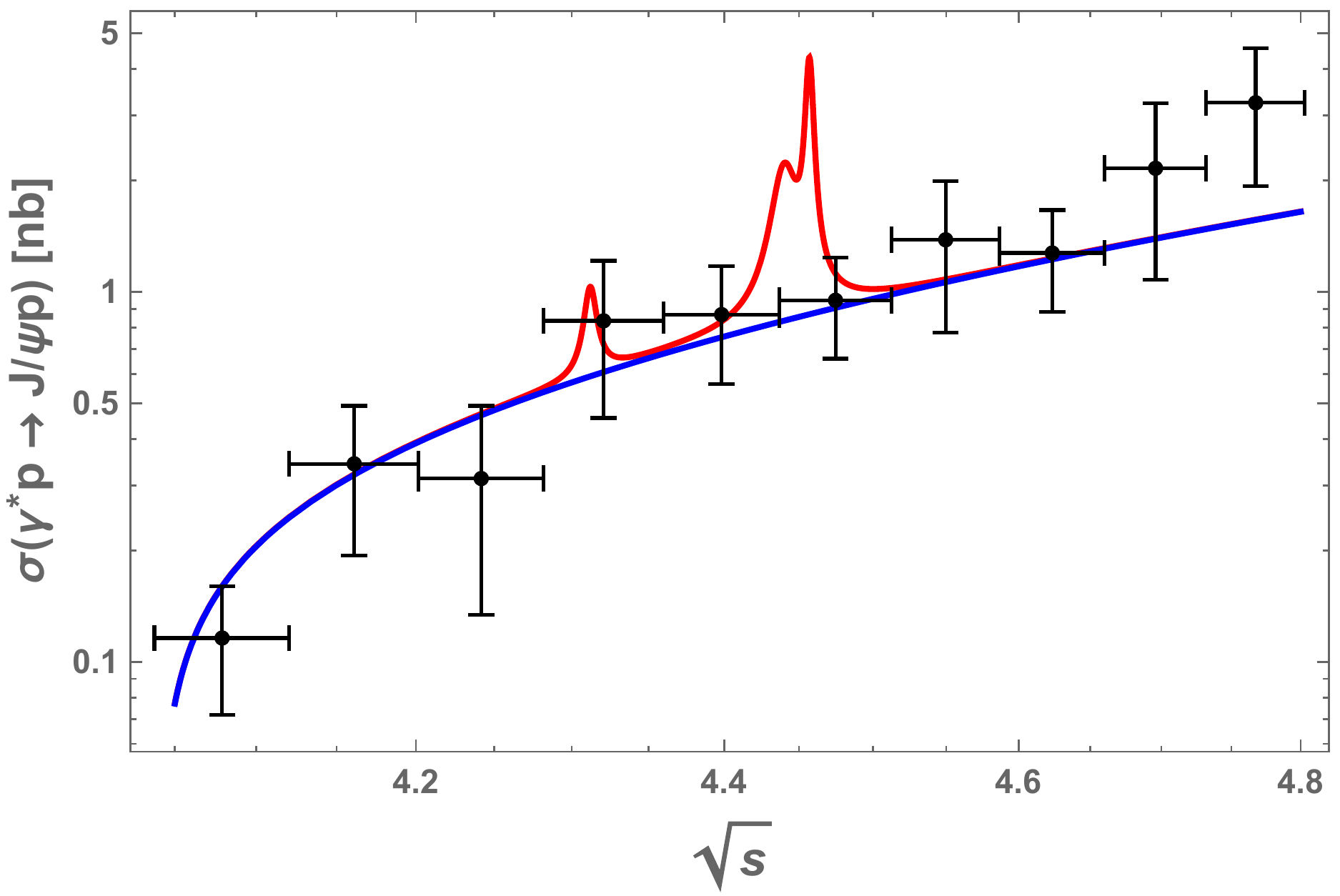}
  \caption{Total cross section for $V=J/\Psi$ photo-production: the blue-solid curve is the t-channel contribution from~\cite{Mamo:2019mka}, the red-solid curve
  is the sum of t- and s-channel contribution showing the three holographic pentaquarks times $\mathcal{N}_s=2.0\times 10^{6}$ to make them visible, and the data are from GlueX~\cite{GlueX:2019mkq}.}
  \label{fig_totcsSplusT}
\end{figure}

\begin{figure}[!htb]
\includegraphics[height=5.5cm]{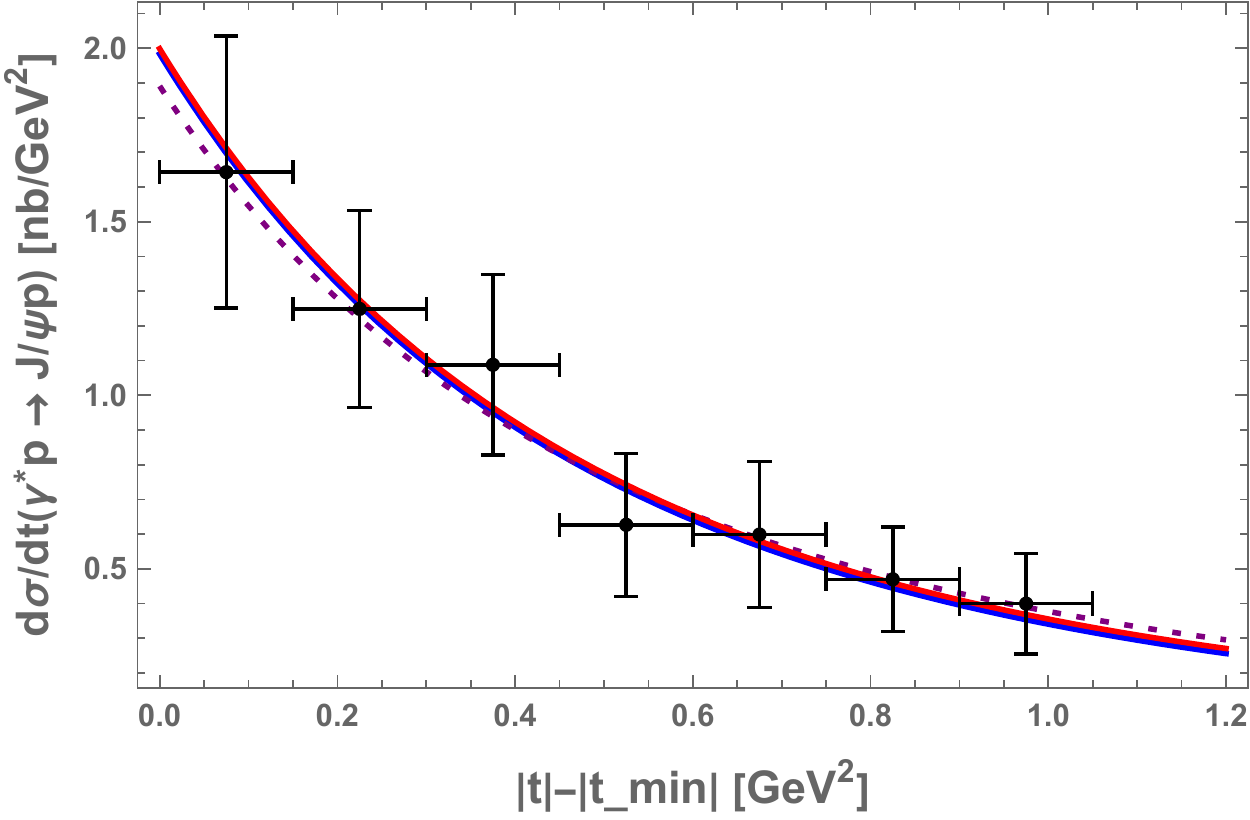}
  \caption{s-channel plus  t-channel  contributions to the photo-production differential cross section  for $V=J/\Psi$, including all three charm pentaquark contributions. The blue-solid curve is the holographic t-channel contribution. The dashed-purple line is from lattice QCD using the holographic kinematic factors. The solid-red line is the total s- and t-channel contribution with the s-channel contribution multiplied by $\mathcal{N}_s=2.0\times 10^{7}$ to make it slightly visible. The data are from GlueX~\cite{GlueX:2019mkq} at $\sqrt{s}=4.6\,\rm{GeV}$.}
  \label{fig_dcsSplusT}
\end{figure}

\section{Analysis}~\label{NUMERICS}

The combined contributions to the differential cross section for the photo-production of a heavy meson $V=J/\Psi, \Upsilon$ is

\bea
\label{DIFFGX}
\left(\frac{d\sigma}{dt}\right)\approx \sum_X \left(\frac{d\sigma}{dt}\right)_X+\left(\frac{d\sigma}{dt}\right)_G
\eea
when restricted to the lowest pentaquark states $[\frac 12\frac 12^-]_{S=0,1}$ and $[\frac 12\frac 32^-]_{S=1}$ contributions.

In Fig.~\ref{fig_totcsS} we show the s-channel contribution to the total cross section for $V=J/\Psi$ photo-production
versus $\sqrt{s}$ in the threshold region, from the three pentaquarks. The smallness of the cross sections follow from the smallness of the Pauli couplings,
as inferred from the holographic construction. Recall that these couplings are commensurate with the bounds on the branching ratios~\cite{GlueX:2019mkq,Cao:2019kst}
as we noted earlier.

The full photo-production cross section with s- and t-channel
contributions combined is shown in Fig.~\ref{fig_totcsSplusT}. The blue-solid curve is the t-channel holographic contribution from~\cite{Mamo:2019mka},
the red-solid curve is our three holographic pentaquark contributions after rescaling by $\mathcal{N}_s=2.0\times 10^{6}$ to make them visible. The data are
from the recent results by the GlueX collaboration~\cite{GlueX:2019mkq}. The smallness of the s-channel pentaquark contribution makes it equally hard to
detect in the differential cross section $d\sigma/dt$ versus $|t|$ as we show in Fig.~\ref{fig_dcsSplusT}. The suppression is larger for the s-channel photo-excitation of
bottom pentaquarks, with the extra penalty

\be
\bigg(\frac{\eta_{P_b}}{\eta_{P_c}}\bigg)^4=\bigg(\frac{\mu_{P_b}}{\mu_{P_c}}\bigg)^4=\bigg(\frac{m_{P_c}}{m_{P_b}}\bigg)^6\approx
\bigg(\frac{4440}{11163}\bigg)^6\approx 0.004
\ee
following from (\ref{MUX}). The t-channel contribution remains unchanged modulo kinematics.

\section{Conclusions}~\label{CONCLUSION}

The holographic construction predicts three holographic pentaquarks with the spin-isospin assignments $[\frac 12\frac 12^-]_{S=0,1}$ and $[\frac 12\frac 32^-]_{S=1}$,
which are split mostly by spin-orbit effects. It also allows for the specific evaluation of their partial decay modes with open or hidden heavy charm (bottom) most of
which are yet to be measured. The holographic pentaquarks are bulk heavy-light meson fields bound to an instanton core. They are dual to a baryon core binding
heavy-light mesons at the boundary with manifest chiral and heavy quark symmetry.

Given the specific spin-isospin assignments for the pentaquark states, the holographic construction allows for an explicit
evaluation of their partial and total strong decay widths with open or hidden charm (bottom) in the final states. This allows
for the extraction of the electromagnetic transition couplings from a proton  to a pentaquark, and decay from a pentaquark.
It also  allows for the evaluation of the  s-channel contribution to photo-production of charmonium (bottomonium) in the threshold
region. In  the holographic construction, the process is mostly dominated by the t-channel graviton in bulk which is dual to
a spin-2 glueball at the boundary~\cite{Mamo:2019mka}.

Our results show that the t-channel contribution at threshold dwarfs the s-channel contribution by several orders of magnitude, making the current search
of the charm (bottom) pentaquark states at current electron facilities out of reach.
This observation appears consistent with the newly reported result  by the GlueX collaboration~\cite{GlueX:2019mkq},
 regarding  the absence of a pentaquark resonance in the photo-production of charmonium.

\vskip 1cm
{\bf Acknowledgements}

I.Z. would like to thank Zein-Eddine Meziani for discussions.
This work is supported by the Office of Science, U.S. Department of Energy under Contract No. DE-FG-88ER40388,
and  by the Polish National Science Centre (NCN) Grant UMO-2017/27/B/ST2/01139. K.M. is supported by the U.S.~Department of Energy, Office of Science, Office of Nuclear Physics, contract no.~DE-AC02-06CH11357, and an LDRD initiative at Argonne National Laboratory under Project~No.~2020-0020.

\section*{Appendix:  \\$X\rightarrow V+p$ decay in the soft wall model}

For completeness, we show that  the vector-like interaction (\ref{REL})  in the (non-relativistic) heavy mass limit, yields a
 decay width $X\rightarrow V+p$ in the soft wall construction,   that is compatible with the one derived
 in the Sakai-Sugimoto model using the bound state approach~\cite{Liu:2021tpq,Liu:2021ixf}. For that, we note that the
 corresponding form factor
(\ref{WMN2}) gives for the partial decay width

\begin{align}
\Gamma_{XV}=\frac{|\vec{q}_V|}{4\pi m_X^2}\frac{\sum_{S,S'}|{\cal W}(-m_V^2)|^2}{2S+1} \ ,
\end{align}
with $\vec{q}_V$ the 3-momentum of the emitted vector meson, and where
$S,S'$ refer to the spin of the initial and final states. In the heavy-quark mass and heavy-nucleon mass limit, the Pauli form factor reduces to

\begin{align}
\label{SPIN}
\frac{\alpha_X}{m_X}\bar u(P,S')(\slashed{q}\gamma^{\mu}-\gamma^{\mu}\slashed{q})u(P,S)
\rightarrow \frac{\alpha_X}{m_X}\bigg(0,\frac{4i}{\sqrt{m_Xm_N}}\,\vec{q}\times \overline{u}_{S^\prime}\vec{\sigma}u_{S}\bigg)
\end{align}
The spin source  in (\ref{SPIN}) is similar to  the spin source in (\ref{1}), although the latter follows by crossing from the former, hence the
difference in the intrinsic parity assignment of the transition vertex.  The corresponding partial decay width is

\begin{align}
\label{DEC1}
 \Gamma_{XV}=|\vec{q}_V|\times \frac{|\vec{q}_V|^2}{8\pi m_X^2}\frac{(8\alpha_X)^2 m_N}{m_X}|{\cal I}(n_X, -m_V^2)|^2
\end{align}
which is to be compared  with the partial decay width we obtained in the Sakai-Sugimoto construction~\cite{Liu:2021ixf}

\begin{align}
\Gamma_{XV}=|\vec{q}_V|\times \frac{ |\vec{q}_V|^2}{8\pi m_X^2} \frac{\lambda^2 m_N}{m_X} \bigg(\frac{M_{KK}}{m_X}\frac{\langle\varphi_V(0)\rangle^2}{\kappa}\bigg) \ .
\end{align}
with $V=n=1$ for the lowest vector mode in bulk. We note that the Pauli-like coupling is the only minimal derivative
term that gives a partial decay width $\Gamma\sim |\vec q_V|^3$ in the non-relativistic limit.
In the heavy quark limit, we can fix  $\alpha_X$ by matching the partial widths

\begin{align}
\label{CONSTANT}
\alpha_X=\lambda\,\bigg(\frac{\langle\varphi_V(0)\rangle}{ 8\sqrt\kappa|{\cal I}(n_X, -m_V^2)|}\bigg)\, \bigg(\frac{M_{KK}}{m_X}\bigg)^{\frac 12}=
\lambda\,C_V\, \bigg(\frac{m_N}{m_X}\bigg)^{\frac 12}\,.
\end{align}

\bibliography{HL}

\end{document}